# 160 MeV laser-accelerated protons from $CH_2$ nano-targets for proton cancer therapy

B. M. Hegelich[1,2], D. Jung[2,3,*], B. J. Albright[1], M. Cheung[3], B. Dromey[3], D. C. Gautier[1], C. Hamilton[1], S. Letzring[1], R. Munchhausen[1], S. Palaniyappan[1], R. Shah[1], H.-C. Wu[4], L. Yin[1], and J. C. Fernández[1]

[1] University of Texas at Austin, Austin, TX 78712, USA

[2] Los Alamos National Laboratory, Los Alamos, New Mexico, 87545, USA

[3] Queen's University Belfast, BT7 1NN, UK

[4] Institute for Fusion Theory and Simulation and Department of Physics, Zhejiang University, Hangzhou 310027, China

[*] B.M.H. and D. J. are joint first authors

**Proton (and ion) cancer therapy has proven to be an extremely effective even superior method of treatment for some tumors [1-4]. A major problem, however, lies in the cost of the particle accelerator facilities; high procurement costs severely limit the availability of ion radiation therapy, with only ~26 centers worldwide. Moreover, high operating costs often prevent economic operation without state subsidies and have led to a shutdown of existing facilities [5,6]. Laser-accelerated proton and ion beams have long been thought of as a way out of this dilemma, with the potential to provide the required ion beams at lower cost and smaller facility footprint [7-14]. The biggest challenge has been the achievement of sufficient particle energy for therapy, in the 150-250 MeV range for protons [15,16]. For the last decade, the maximum experimentally observed energy of laser-accelerated protons has remained at ~60 MeV [17]. Here we the experimental demonstration of laser-accelerated protons to energies**

**exceeding 150 MeV, reaching the therapy window. This was achieved through a different acceleration regime rather than a larger laser, specifically a 150 TW laser with $CH_2$ nano-targets in the relativistically transparent regime [18,19]. We also demonstrate a clear scaling law with laser intensity based on analytical theory, computer simulations and experimental validation that will enable design of a prototype system spanning the full range of therapeutically desirable energies.**

A number of challenges have been identified in realizing a laser-based proton therapy system, among them the large energy spread of the laser-accelerated proton beam, beam transport issues, and secondary radiation in the form of gamma rays from the laser-plasma interaction or neutrons from beam-wall interactions. However, the largest challenge relates to a prerequisite for practical utility: the insufficient proton energies demonstrated from laser-plasma accelerators: ~60 MeV, only 1/3-1/4 of what is needed. Here, for the first time, we demonstrate the acceleration of protons to energy in excess of 150 MeV, which access the useful energy range for therapy applications as well as to study other challenges like beam transport design, dose rate planning, etc. Furthermore, we were able to demonstrate these energies with the rather modest Trident laser system of only ~150 TW power, roughly 1/6 of the power available e.g. at the new Bella laser at LBNL or the Texas Petawatt laser at UT Austin. This breakthrough is enabled by a paradigm shift in the laser-target interaction. Rather than relying on surface acceleration in the so-called Target Normal Sheath Acceleration (TNSA) regime, as practised for the last decade, we succeeded in reaching a regime of relativistic transparency [18], where a still overdense (and thus opaque) target is rendered transparent by the lowering of the plasma frequency resulting from the laser driving the whole electron population to relativistic energies. The transition of a target to a state of relativistic transparency allows the laser to propagate through the target at high density and interact with all particles in the focal volume, thus increasing the energy coupling of laser-light to particles and the overall efficiency of the acceleration both in particle energy and number [20]. This mechanism, called "Break-Out Afterburner" (BOA) acceleration, has been predicted in simulations [19,21] and demonstrated for Carbon ions [22-24] and Deuterons [25]. Here we optimized the target for the acceleration of protons, allowing the realization of a threefold energy increase over previous numbers using the same laser in the TNSA regime[26].

The experiments were conducted at the Trident laser facility at Los Alamos National Laboratory [27]. Trident is a high-power 150TW glass laser and delivers 80 J in ~550 fs at a wavelength of 1054 nm. As final focusing optic, off-axis parabolic mirrors with different free apertures (F/8, F/3 and F/1.5) focus the beam to spot sizes of 8μm, 3.5μm, and 1.5μm, respectively, containing ~60% of the energy in this first maximum. This yields intensities from $8 \times 10^{19}$ W/cm$^2$ to $2 \times 10^{21}$ W/cm$^2$ at constant pulse length and energy. Trident's exceptionally high laser contrast (<$10^{-9}$ at 50ps, <$10^{-6}$ at 5ps and $10^{-4}$ at 2ps before the peak of the pulse) ensured interaction of the peak laser pulse with a highly over-dense plasma[18], even for targets with thickness of only a few 100 nm. The laser was used to irradiate synthetically made $CH_2$ plastic foil targets (see Methods) with thicknesses of 200 nm to 800 nm. Angularly resolved proton spectra were measured using an ion wide angle spectrometer (iWASP)[28] with an energy resolution of better than 10% at 200MeV.

In the experiment, the highest proton energy observed with the F/3 focusing setup (peak intensity ≈ $4 \times 10^{20}$ W/cm$^2$) was obtained from nominally 240 nm thick $CH_2$ targets. An example of such a spectra, measured with the iWASP over an angle of ~ -1° to ~20° (0° being the laser and target normal direction) is shown in Figure 1. The maximum proton energies are found off-axis from 8° to 14°, consistent with expectations for BOA acceleration with tightly-focused laser beams, as predicted by simulations [29] and recently measured [25]. The maximum energy is well in excess of 100 MeV, about twice the value from NOVA [17], and thus the first substantial increase reported in a decade.

The image plate detector used in the iWASP returns a signal in PSL units (Photo Stimulated Luminecence). In order to relate PSL to absolute particle numbers we used two published calibrations for protons on image plates [30,31]. These two calibrations represent the upper and lower end of the error bars in the spectra shown in Fig. 1. For a 100 msr beam this equates to $10^7 - 10^8$ protons with energy >100 MeV, enough particles per shot for medical applications if a laser with repetition rates of 1-10Hz and a suitable target design are used.

To confirm the spectrometer measurements with image plates, we repeated the shots at the optimal target thickness of 200-300nm, replacing the iWASP with a stack of three CR-39 nuclear track detectors sandwiched between 2.5mm Cu and 3.3mm Ta moderators. The exact stack geometry is described in in the supplemental material, the three data points at 39 MeV, 75 MeV, and 100 MeV are shown in Figure 1 and are in agreement with the spectrometer data.

To further increase the proton energies, as well as to test the scaling with laser intensity and the predictive capability of our models, we conducted an experiment where we changed from F/3 focusing to an F/1.5 off-axis parabolic mirror (OAP). This results in a 1.5 micron radius focal spot, increasing the intensity to $\approx 1.6 \times 10^{21}$ W/cm$^2$ at the same laser energy and pulse duration. As is shown in Figure 2, the maximum observed proton energy for the F/1.5 geometry exceeds 160 MeV (160 MeV represents the useful upper energy resolution of the current iWASP spectrometer, limited by magnet strength deflection path length and thickness of the W-entrance aperture). This result represents a further significant increase that is in good qualitative and quantitative agreement with both our analytic model and the kinetic simulations using the VPIC code [32]. The two-dimensional (2D) VPIC simulation for the shot parameters predicts 200 MeV protons, as is also indirectly suggested by neutron measurements performed in a later campaign [25]. A 2D VPIC simulation for an F/1 Trident focusing scenario published earlier[33] predicts proton energies of up to 250 MeV for an intensity of $5 \times 10^{21}$ W/cm$^2$.

Our experiments apply the BOA technique for laser-ion acceleration [19,22,29], one of several next-generation ion-acceleration methods that have been developed in recent years[20,21,23,33-38]. In the BOA, a high-contrast, high-intensity laser pulse interacts with a nanometer-scale target, heating the electrons beneath the laser spot. As the target heats, it expands, becoming relativistically transparent at a time $t_1$ when the electron density in the target $n_e \sim n_{cr}/\gamma$, where $n_{cr} = m_e \omega_0^2 / 4\pi e^2$ is the classical critical density for a laser of frequency $\omega_0$ and $\gamma$ is the mean Lorentz factor of the target electrons, which have mass $m_e$ and charge e. The transition to relativistic transparency begins a period of enhanced coupling of laser momentum to target electrons and, through the generation of self-consistent electrostatic fields, target ions[39,40]. Significant ion acceleration proceeds within the target until time $t_2$, when target expansion has proceeded to the point where the plasma is classically underdense, with diminished coupling between the laser pulse and electrons. Optimal acceleration with the BOA has been found to occur when the arrival at the target of the peak intensity of the laser pulse is between $t_1$ and $t_2$ and the acceleration scales as $a_0 =$

$eE_0/m_e\omega_0c$, the scaled laser electric field, two features that distinguish the BOA from other mechanisms [23]

High-resolution, 2D, particle-in-cell simulations using the VPIC code confirm these dynamics. In our simulations, shown in Fig. 3., a thin foil of solid-density, $CH_2$ plasma is initialized, upon which a normally incident, intense laser pulse is directed. In contrast with prior simulations of solid-density carbon targets with trace fractions of protons (targets that "self-clean" during $t<t_1$ when the target is still opaque to the laser[21]), in simulations of proton-rich targets, a sizeable population of protons remains within the main target layer at $t_1$. This allows the peak accelerating electric fields to act on these protons, thus accelerating them to the higher observed proton energies.

We note that other laser-ion acceleration techniques such as Radiation Pressure Acceleration (RPA) [35-37] have been reported to possess more favourable scaling with $a_0$ than the BOA, thus enabling the possibility of higher peak ion energy for given laser intensity. Practical difficulties associated with achieving RPA on our laser system prevented our use of RPA acceleration in this experiment, however. Nevertheless, if realizable, an RPA-based proton acceleration system could have significant advantages over BOA-based systems.

*Scaling*

Figure 4 shows a plot of the experimentally obtained maximum protons energies for three different laser intensities (red stars) achieved with different final focusing optics (off-axis parabolic mirrors with free aperture of F/8, F/3 and F/1.5). Intensities are plotted as normalized laser amplitude. The data points are in remarkably good agreement with results from VPIC simulations (blue squares) of the experimental target and laser parameters and indicate a linear correlation between the laser intensity and the maximum proton energy (for fixed laser energy and duration), as predicted by a prior analytic study[40]. The linear scaling can be seen in the fit of the experimental data (grey solid line), which gives a correlation of 4.5MeV per unit laser amplitude plus 29.3 MeV for maximum proton energies.

## *Conclusion and Outlook*

With the first demonstration of medically relevant proton energies from a laser-accelerator, a major hurdle to laser-driven systems has been removed. The demonstrated 160 MeV protons are already sufficiently energetic for wide range of applications, having a range of 17.4 cm in water. Furthermore, given the scaling behaviour we have demonstrated, it is straightforward to extrapolate to a laser system capable of generating the full 250 MeV. Even using a conservative estimate, assuming no improvements due to advanced targets, an existing system like the Texas Petawatt could, in principle, generate ion beams of 250 MeV with $CH_2$ targets or up to almost 500 MeV with cryogenic hydrogen targets[23] once the required laser pulse contrast is achieved. We note that all existing systems are based on 30-40 year old technology, severely limiting the available repetition rate. However, even today, various designs, both flash lamp and diode based, exist to build a sufficiently powerful laser with 1-10 Hz repetition rate. A flash-lamp, liquid-cooled amplifier prototype developed by the University of Texas and National Energetics in Austin has already demonstrated 0.1 Hz operation and a 3 Hz version capable of producing 100-200J is being developed. Such a laser system could be fielded for ~$10M, making a prototype facility for medical research at relevant parameters feasible.

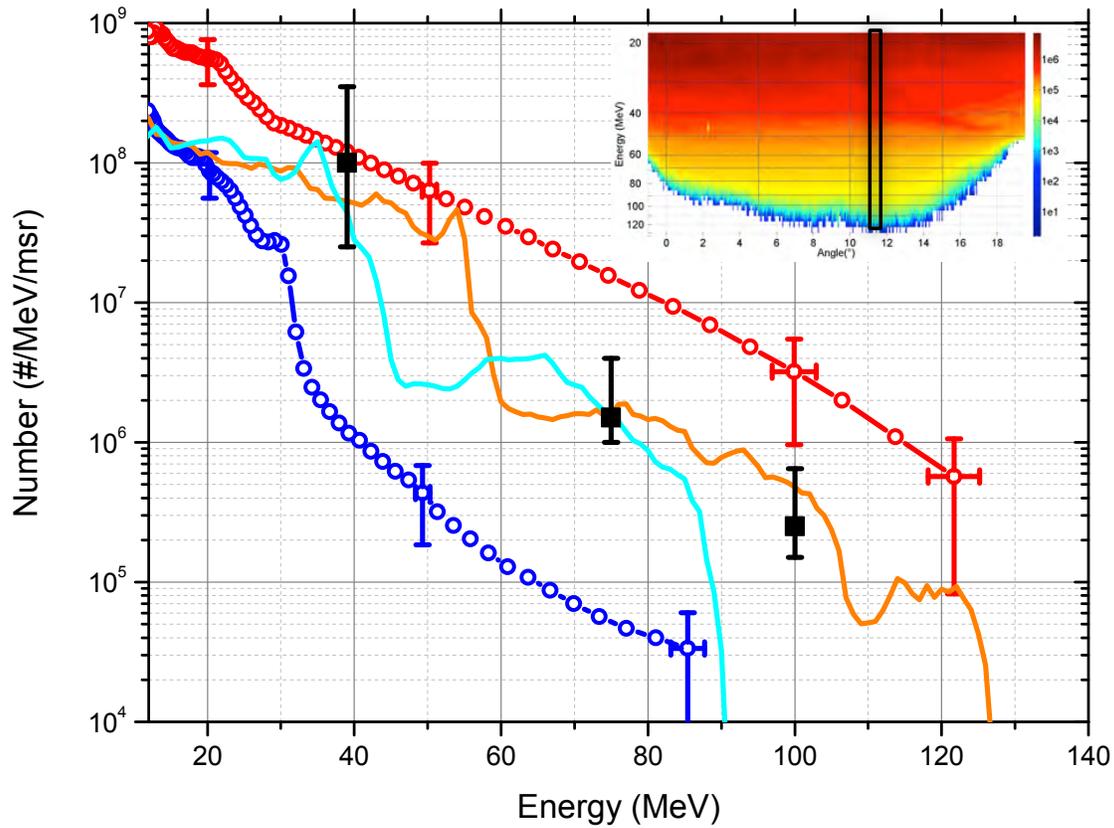

Figure 1: Proton spectra from 240nm (red circles) and 300nm (blue circles) thin $CH_2$ plastic targets from interaction with a ~80J, ~600fs, Trident laser pulse with peak intensity $\approx 4 \times 10^{20}$ W/cm$^2$. The spectra are line-outs at 10 degrees from a wide angle measurement from 0-25 degree (full spectrum shown in inset). The black square are protons measured by a stack of CR-39 interspaced with W-filters. Also shown are spectra from 2D VPIC simulations for Trident conditions and a 240nm (orange) and 300nm (cyan) $CH_2$ target. The simulations do not include the effects of the finite laser pre-pulse.

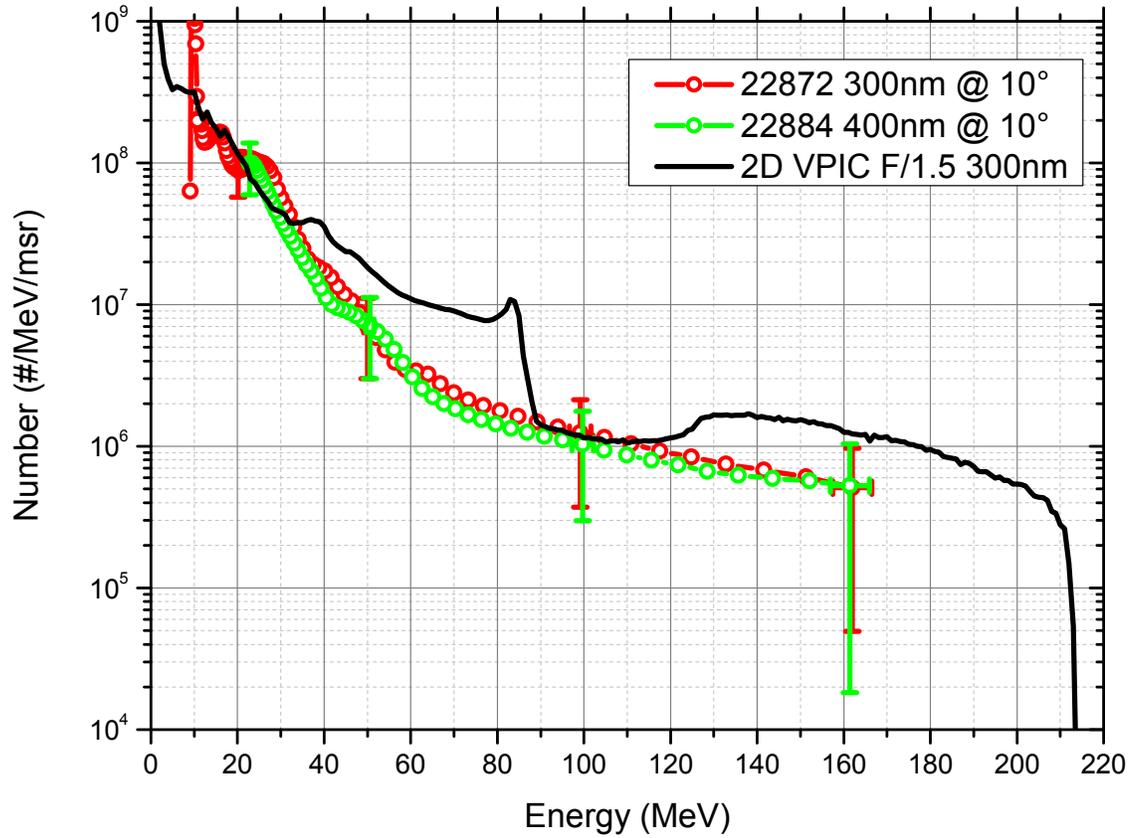

**Figure 2**: Observed proton spectra using an F/1.5 OAP, resulting in an on-target intensity of ≈1.6x10$^{21}$ W/cm$^2$, in conjunction with a 300-400nm CH$_2$ target. The current spectrometer is only accurate to proton energies of ~160 MeV. PIC simulations and neutron measurements indicate proton energies beyond 200 MeV.

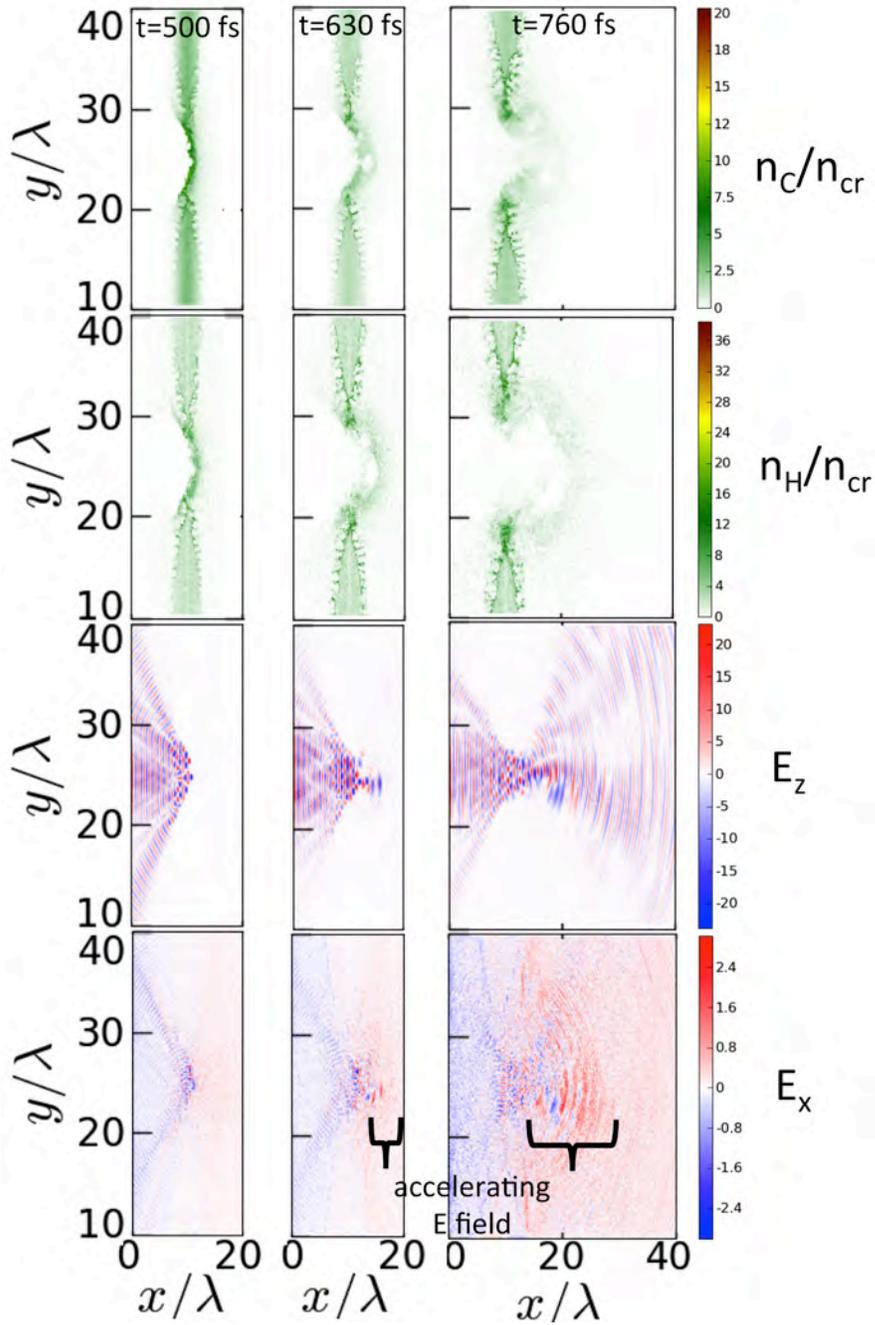

Figure 3: Two-dimensional VPIC simulations of the Trident laser pulse interacting with a solid-density $CH_2$ target undergoing a transition to relativistic induced transparency. Shown are data near the laser focal spot (the actual volume simulated is much larger than shown) at three instants in time: 500 fs, (just before transparency), 630 fs (during transparency, onset of BOA), and 760 fs (well into the BOA acceleration regime). The top two rows show carbon and hydrogen charge density, respectively, normalized to the critical density. The third row is the laser field $E_z$ normalized to $a_0 = mc\omega_L/e$.

The fourth row is the longitudinal accelerating electric field $E_x$, normalized to $a_0$. The largest longitudinal fields occur after transparency. Because of the abundance of protons in the target, the protons are not fully expelled from the accelerating volume prior to transparency and as indicated, sizable $E_x$ field overlap the proton density.

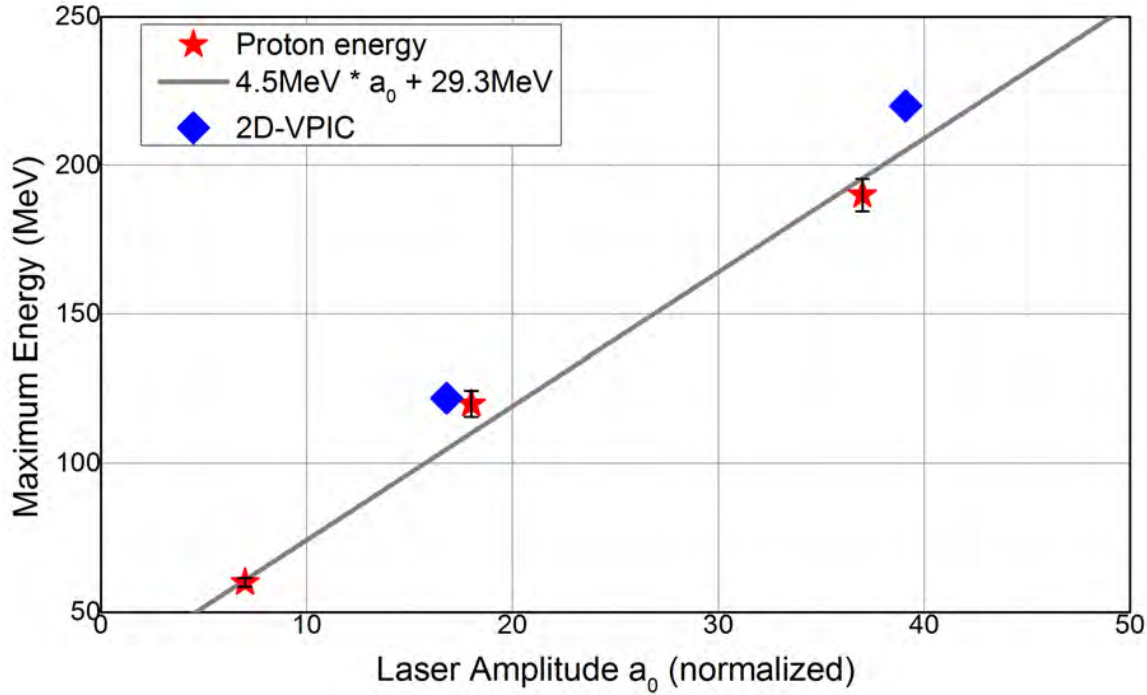

Figure 4: proton energy scaling with $a_0$, measured on Trident for constant laser energy and pulse duration, using 3 different off-axis parabolas to ensure different focal spot sizes at optimal focusing conditions. Experimental proton energies scale linearly with $a_0$, which is consistent with analytic models [40,41] and simulations [21,29,33,42].

Methods

Laser system:

The Trident laser facility has a short pulse beam with 80 J in ~500 fs (best optimized performance) at a wavelength of 1054 nm. The laser has s-polarization and is focused with an off-axis parabolic mirror (OAP). During this experiment three different OAPs with free aperture F/8, F/3 and F/1.5 have been used. The F/8 OAP yields a measured on-target focus of 11.6μm radius (1/e2-condition, containing > 60% of the laser energy) and a peak intensity $8.4\times10^{19}$ W/cm$^2$; the F/3 OAP yields a measured on-target focus of 6.2μm FWHM and a peak intensity $4\times10^{20}$ W/cm$^2$; the F/1.5 OAP yields a measured on-target focus of ~2.5μm radius and a peak intensity $1.6\times10^{21}$ W/cm$^2$. Exceptionally high laser contrast ($10^{-7}$ at -4 ps) enabled by the OPAPE technique[43] ensures interaction of the laser pulse with a highly over-dense plasma[18] for targets in the nm-thickness range.

Targets:

The nano-CH$_2$ targets were prepared by spin-casting from dilute polymer solutions as follows: polymethylpentene (TPX, medium molecular weight, Sigma-Aldrich) was added to cyclohexane with stirring and gentle heating, dissolution was typically complete within 1 - 2 hours. Concentrations were prepared between 1.4 and 1.9 wt% depending on final desired film thickness. The warm polymer solution (0.5 mL) was dropped onto a glass slide spun at 3500 rpm for 30 seconds. The targets were characterized for thickness by stylus profilometry. Films were mounted onto silicon target stalks as follows: immersion of the coated glass slides in water allowed the polymer to be floated off onto the surface of the water. The films were then mounted onto target stalks by bringing the stalks underneath the water and scooping the films out. No adhesives were necessary to hold films in place. Films were prepared from ~200 nm to 1 um in thickness.

Diagnostics:

The main ion-beam diagnostic was the ion Wide Angle Spectrometer (iWASP), capable of simultaneously measuring proton and carbon spectra over a 25 degree angle[28]. At the expected maximum ion energy of 150 MeV protons, the energy resolution is ~9%. Both CR-39 nuclear track detectors and image plates were used as detectors. However protons with energy >10 MeV when passing through the CR39 will not leave tracks in it. Therefore, we used, image plates as the primary detector for proton spectra. The image plate detector used in the iWASP returns a signal in PSL units. In order to relate that to absolute particle numbers we used two published calibrations for protons on image plates[30,31]. These calibrations bracket the measured particle spectrum, yielding particle numbers of ~$10^7 – 10^8$ protons with energy >100 MeV, respectively.

To confirm our image-plate measurements, additional shots were performed where the iWASP was replaced by a stack of CR-39 and Ta moderators, to obtain measurements on CR-39 at 3 different proton energies. The stack has a transverse size of 5x10cm and was placed 3cm behind the target. The individual layers in the stack detected 39 MeV, 75 MeV, and 100 MeV protons respectively, with an energy width of the order of 1MeV. In the case of carbon ions the corresponding energies are 880 MeV, 1.67 GeV and 2.29 GeV. The CR-39 was analyzed using our ELBEK automated scanning system. All three layers show a clear proton signal on both sides, the counted numbers for each layer are within an order of magnitude of both IP calibrations if one takes into account the full solid beam angle of 50-100msr. About 0.1-1% of the tracks were observed to be significantly larger and darker than the majority of tracks, originating from either tracks caused by heavier ions (Carbon) or proton knock-ons giving energy to carbon or oxygen ions in the CR-39 or even Cu or Ta ions at the interface to the moderator.

Simulations:

The particle-in-cell simulations employed the explicit, relativistic, particle-in-cell simulation code VPIC in 2D Cartesian geometry. The simulation domain was of size 100 microns (longitudinally) by 45 microns (transversely) and employed absorbing particle and field boundary conditions. The laser was modelled as a Gaussian beam (spatially) with peak $a_0$=13, diffraction-limited spot size 7 microns, and a sine-squared $E_0$ envelope with FWHM 540 fs. The target was placed 10 microns from the left boundary and had initial thickness 300 nm and electron density $n_{e0}$=280 $n_{cr}$. The plasma was modelled with 250 particle/cell/species and initial temperatures $T_e$=100 keV and $T_i$=10eV. The simulation domain was 50000 by 9000 cells, resolving both plasma skin depth and initial Debye length. The simulations were run until time t=1.37 ps.